\documentclass{article}

\usepackage{amscd,amsmath,amssymb}

\newcommand{\mf}{\mathbf{f}}

\newcommand{\mV}{\mathbf{V}}
\newcommand{\mbV}{\overline{\mathbf{V}}}
\newcommand{\mpsi}{\boldsymbol{\psi}}
\newcommand{\mphi}{\boldsymbol{\phi}}
\newcommand{\mbpsi}{\bar{\boldsymbol{\psi}}{}}

\def\DH{\rm I\kern-1.5pt\rm H\kern-1.5pt\rm I}

\def\DR{\rm I\kern-1.45pt\rm R}
\def\DC{\kern2pt {\hbox{\sqi I}}\kern-4.2pt\rm C}

\newcommand{\cE}{{\cal E}{}}
\newcommand{\cD}{{\cal D}{}}
\newcommand{\cA}{{\cal A}}

\newcommand{\bF}{{\overline F}}
\newcommand{\bD}{{\overline D}{}}
\newcommand{\bV}{{\overline V}{}}
\newcommand{\bnabla}{{\overline \nabla}{}}

\newcommand{\bW}{{\overline W}}
\newcommand{\bQ}{{\overline Q}{}}
\newcommand{\bB}{{\overline B}{}}
\newcommand{\bS}{{\overline S}{}}

\newcommand{\bpsi}{{\bar\psi}}

\newcommand{\nn}{\nonumber}
\newcommand{\ba}{\begin{array}}
\newcommand{\ea}{\end{array}}
\newcommand{\be}{\begin{equation}}
\newcommand{\ee}{\end{equation}}
\newcommand{\bea}{\begin{eqnarray}}
\newcommand{\eea}{\end{eqnarray}}
\newcommand{\bi}{\begin{itemize}}
\newcommand{\ei}{\end{itemize}}

\newcommand{\p}[1]{(\ref{#1})}
\newcommand{\mD}{\mathbb{D}{}}
\newcommand{\mbD}{\overline{\mathbb{D}}{}}

\def\im{{\rm i}}

\begin{document}\thispagestyle{empty}

\begin{flushright}
\end{flushright}\vspace{2cm}

\begin{center}
{\Large\bf Space-filling D3-brane within coset approach}
\end{center}
\vspace{1cm}

\begin{center}
{\large\bf S.~Bellucci${}^a$, N.~Kozyrev${}^{b}$, S.~Krivonos${}^{b}$, A.~Sutulin${}^{a,b}$,
}
\end{center}

\begin{center}
${}^a$ {\it
INFN-Laboratori Nazionali di Frascati,
Via E. Fermi 40, 00044 Frascati, Italy} \vspace{0.2cm}

${}^b$ {\it
Bogoliubov  Laboratory of Theoretical Physics, JINR,
141980 Dubna, Russia} \vspace{0.5cm} \\

{\tt bellucci@lnf.infn.it, nkozyrev, krivonos, sutulin@theor.jinr.ru}

\end{center}
\vspace{2cm}

\begin{abstract}\noindent
We derive the component on-shell action of the space-filling D3-brane, {\it i.e.} $N=1$ supersymmetric Born-Infeld action, within the nonlinear
realization approach. The covariant Bianchi identity defining the $N=1$, $d=4$ vector supermultiplet has been constructed by introducing a new bosonic
Goldstone superfield associated with the generator of the $U(1)$ group, which transforms to each other the spinor generators of unbroken
and spontaneously broken $N=1$, $d=4$ supersymmetries. The first component of this Goldstone superfield is the auxiliary field of the vector supermultiplet
and, therefore, the Bianchi identity can be properly defined. The component action of the D3-brane has a very simple form, being written in terms
of derivatives covariant with respect to spontaneously broken supersymmetry - it just mimics its bosonic counterpart.
\end{abstract}

\newpage
\setcounter{page}{1}
\setcounter{equation}{0}
\section{Introduction}
The idea of partial breaking of global supersymmetry is quite useful being applied to superbranes (see e.g. \cite{1} and references therein). In such an approach
the physical worldvolume superbrane  degrees of freedom are described by  Goldstone superfields. A part of the supersymmetry is realized linearly providing
the standard worldvolume superfield description of the superbrane, while the rest of the target supersymmetry is realized nonlinearly. The main difficulties encountered in the application of the standard methods of nonlinear realization \cite{NLR1,NLR2} to the construction of the superbrane actions is the absence of the systematic
procedure for deriving the superfield actions. This is the result of partial breaking of the supersymmetry, because in the case of the total supersymmetry
breaking \cite{VA} the action can be immediately constructed in terms of the corresponding Cartan forms. The reason for this is obvious: the superfield
Lagrangian is not invariant with respect to supersymmetry. Instead, it is shifted by the full space-time derivative under supersymmetry transformations and, therefore,
it cannot be constructed from the Cartan forms.

It has been shown in \cite{BIK1} that the nonlinear realization approach can be used for deriving the superfield constraints and the equations of motion as the
direct covariantization of those describing the free system. However, the situation with obtaining the action at the superfield and/or at the component level was still unsolved.
As a partial way out, it was proposed in \cite{BKS1,BKS2} to construct the on-shell component action within the nonlinear realization approach. The main novelty
of the proposed scheme was shifting attention from unbroken to broken supersymmetry. Indeed, it is always possible to realize the spontaneously broken part of
supersymmetry in such a way as to keep the $\theta$-coordinates of unbroken supersymmetry invariant. If such a basis is chosen, then each component of Goldstone
superfields transforms independently from the remaining ones. Then, one may use the broken supersymmetry to construct the covariant measure and the covariant derivatives
with respect to it, using them to derive the component action. Within such an approach the component actions for many interesting branes were constructed
\cite{ourbranes1,ourbranes2,ourbranes3}. However, in all considered cases the physical bosonic degrees of freedom contain only scalar fields. These scalars
were associated with the Goldstone fields for partially broken translations and, therefore, all physical components appeared as the coordinates of the corresponding
coset. Thus, all D-branes, including supersymmetric Born-Infeld theories were out of the game. The main difficulty preventing from the direct extension of our scheme to the
cases of theories with vector fields was the impossibility to derive the covariant Bianchi identity. So, we meet the same problem that has been discussed  in the pioneering paper on this subject \cite{BG2}. Roughly speaking, the direct covariantization of the constraints defining
the $N=1, d=4$ vector supermultiplet, i.e.
$$
D_\alpha \mpsi^\alpha+ \bD{}^{\dot\alpha} \mbpsi{}_{\dot\alpha}=0\quad \Rightarrow \quad \nabla_\alpha \mpsi^\alpha+ \bnabla{}^{\dot\alpha} \mbpsi{}_{\dot\alpha}=0
$$
implies the equations of motion. The main task of this paper is to derive the proper covariant Bianchi identity (Section 3) and to construct the corresponding
component action (Section 4).

\setcounter{equation}{0}
\section{Sketch of  Bagger \& Galperin results}
The superfield action of D3-brane, i.e. $N=1$ supersymmetric Born-Infeld action \cite{CF}, realizing the $N=2 \rightarrow N=1$ partial spontaneous breaking
of $d=4$ supersymmetry, was obtained many years ago in the paper of J.~Bagger and A.~Galperin \cite{BG2} within the linear realization approach.
Then, the same action has been reproduced using the nilpotent superfields in \cite{RT}. In any case, the main conclusion which follows from these
 papers is the claim that partially broken $N=2$
supersymmetry with the $N=1$ vector supermultiplet as the Goldstone one uniquely fixes the action to be the $N=1$ supersymmetric Born-Infeld action of \cite{CF}.
Therefore, it is quite natural to construct the D3-brane action within the nonlinear realization approach, at least at the component level.
And indeed the paper \cite{BG2}  started with the corresponding consideration. Unfortunately, the story stops quite quickly, because the proper
covariantization of the irreducibility constraints, which would single out the $N=1$ vector supermultiplet, is not trivial.
Let us remind shortly this part of the paper \cite{BG2} to visualize the problem and
to fix our notations.

Starting from $N=2$, $d=4$ supersymmetry algebra
\be\label{algebra}
\left\{  Q_\alpha, \bQ_{\dot\alpha}   \right\} = 2 P_{\alpha\dot\alpha}, \quad \left\{  S_\alpha, \bS_{\dot\alpha}   \right\} = 2 P_{\alpha\dot\alpha},
\ee
where $Q_\alpha$ and $S_\alpha$ are  the supersymmetry generators and $P_{\alpha\dot\alpha}$ is the generator of four dimensional translations, one may realize this algebra
in the coset parameterized as
\be\label{coset}
g = e^{\im x^{\alpha\dot\alpha} P_{\alpha\dot\alpha}} e^{\im \left( \theta^\alpha Q_\alpha + \bar\theta_{\dot\alpha}\bQ^{\dot\alpha}   \right) } e^{\im \left( \mpsi^\alpha S_\alpha + \mbpsi_{\dot\alpha}\bS^{\dot\alpha}   \right) }.
\ee
Using the Cartan forms
\bea\label{CF}
g^{-1} d g & = & \im \omega_P^{\alpha\dot\alpha} P_{\alpha\dot\alpha} +\im \omega_Q{}^\alpha Q_\alpha+\im \bar\omega_{Q\; \dot\alpha} \bQ^{\dot\alpha}+\im \omega_S{}^\alpha S_\alpha+\im \bar\omega_{S\; \dot\alpha} \bS^{\dot\alpha} , \nn \\
\omega_P^{\alpha\dot\alpha} & = & dx^{\alpha\dot\alpha} -\im \left( \theta^\alpha d \bar\theta{}^{\dot\alpha}+ \bar\theta{}^{\dot\alpha} d \theta^\alpha+
\mpsi^\alpha d \mbpsi{}^{\dot\alpha}+ \mbpsi{}^{\dot\alpha} d \mpsi^\alpha\right), \nn \\
\omega_Q{}^\alpha & = & d\theta^\alpha, \quad \bar\omega_Q{}_{\dot\alpha} = d \bar\theta_{\dot\alpha}, \quad \omega_S{}^\alpha = d\mpsi^\alpha,\quad
\bar\omega_S{}_{\dot\alpha} = d \mbpsi_{\dot\alpha},
\eea
one may define the covariant derivatives
\bea\label{CD}
&& \nabla_{\alpha\dot\alpha} = \big ( E^{-1}\big)_{\alpha\dot\alpha}^{\beta\dot\beta} \; \partial_{\beta\dot\beta}, \quad
E_{\alpha\dot\alpha}^{\beta\dot\beta}=\delta_\alpha^\beta \delta_{\dot\alpha}^{\dot\beta}-\im \mpsi{}^\beta \partial_{\alpha\dot\alpha} \mbpsi{}^{\dot\beta}-
\im \mbpsi{}^{\dot\beta} \partial_{\alpha\dot\alpha} \mpsi{}^{\beta}, \nn \\
&&\nabla_\alpha =D_\alpha-\im \big( \mbpsi{}^{\dot\beta} D_\alpha \mpsi{}^\beta + \mpsi{}^\beta D_\alpha \mbpsi{}^{\dot\beta}\big) \nabla_{\beta\dot\beta}=
D_\alpha-\im \big( \mbpsi{}^{\dot\beta} \nabla_\alpha \mpsi{}^\beta + \mpsi{}^\beta \nabla_\alpha \mbpsi{}^{\dot\beta}\big) \partial_{\beta\dot\beta}\,, \nn \\
&&\bnabla_{\dot\alpha} =\bD_{\dot\alpha}-\im \big( \mbpsi{}^{\dot\beta} \bD_{\dot\alpha} \mpsi{}^\beta + \mpsi{}^\beta \bD_{\dot\alpha} \mbpsi{}^{\dot\beta}\big) \nabla_{\beta\dot\beta}
= \bD_{\dot\alpha}-\im \big( \mbpsi{}^{\dot\beta} \bnabla_{\dot\alpha} \mpsi{}^\beta + \mpsi{}^\beta \bnabla_{\dot\alpha} \mbpsi{}^{\dot\beta}\big) \partial_{\beta\dot\beta}\,.
\eea
The expression of the flat covariant spinor derivatives has the standard definition
\be
D_\alpha=\frac{\partial}{\partial\theta^\alpha} -\im \bar\theta{}^{\dot\alpha} \partial_{\alpha\dot\alpha},\quad
\bD_{\dot\alpha}=-\frac{\partial}{\partial\bar\theta{}^{\dot\alpha}} +\im \theta^{\alpha} \partial_{\alpha\dot\alpha},\quad
\big\{ D_\alpha, \bD_{\dot\alpha}\big\} = 2 \im \partial_{\alpha\dot\alpha}.
\ee
The covariant derivatives given in \p{CD} obey the following (anti)commutation relations
\bea\label{nabladercomms}
\big \{ \nabla_\alpha, \nabla_\beta   \big \} &=& -2\im \big( \nabla_\alpha \mpsi^\gamma \nabla_\beta \mbpsi{}^{\dot\gamma}
+ \nabla_\beta \mpsi^\gamma \nabla_\alpha \mbpsi{}^{\dot\gamma}  \big) \nabla_{\gamma\dot\gamma}, \nn \\
\big \{ \nabla_\alpha, \bnabla_{\dot\alpha}   \big \}
&=& 2\im \nabla_{\alpha\dot\alpha} -2 \im \big( \nabla_\alpha \mpsi^\gamma \bnabla_{\dot\alpha} \mbpsi{}^{\dot\gamma}
+ \bnabla_{\dot\alpha} \mpsi^\gamma \nabla_\alpha \mbpsi{}^{\dot\gamma}  \big)\nabla_{\gamma\dot\gamma}, \nn \\
\big [ \nabla_\alpha, \nabla_{\beta\dot\beta}   \big ]
&=& 2\im \big ( \nabla_\alpha \mpsi^{\gamma}  \nabla_{\beta\dot\beta} \mbpsi{}^{\dot\gamma}
+ \nabla_{\beta\dot\beta} \mpsi^{\gamma} \nabla_\alpha \mbpsi{}^{\dot\gamma}    \big ) \nabla_{\gamma\dot\gamma} , \nn \\
\big[ \nabla_{\alpha\dot\alpha}, \nabla_{\beta\dot\beta}   \big]
&=& 2\im \big ( \nabla_{\alpha\dot\alpha} \mpsi^{\gamma}  \nabla_{\beta\dot\beta} \mbpsi{}^{\dot\gamma}
+ \nabla_{\alpha\dot\alpha} \mbpsi{}^{\dot\gamma} \nabla_{\beta\dot\beta} \mpsi^{\gamma}    \big) \nabla_{\gamma\dot\gamma} .
\eea
The final step is to find the covariant version of the flat constraints
\be\label{free}
\bD_{\dot\alpha} W_\alpha =0, \quad D_\alpha {\overline W}_{\dot\alpha}=0 \quad (a), \qquad D_\alpha W^\alpha+ \bD{}^{\dot\alpha}{\overline W}{}_{\dot\alpha} =0 \quad (b),
\ee
selecting the $N=1, d=4$ vector supermultiplet $W_\alpha, \bW{}_{\dot\alpha}$ \cite{BW}.
It was shown in \cite{BG2} that the chirality constraints (\ref{free}a) can be directly covariantized
\be\label{covchir}
\bnabla_{\dot\alpha} \mpsi_\alpha =0, \quad \nabla_\alpha \mbpsi_{\dot\alpha}=0.
\ee
The reason why these constraints are correct is that they are just the $d\bar\theta_{\dot\alpha}$ and $d \theta^\alpha$ parts
of the Cartan forms $\omega_S{}^\alpha$ and $\bar\omega_S{}_{\dot\alpha}$ \p{CF}, respectively. Remembering that the Cartan forms are
invariant with respect to all transformations of $N=2, d=4$ Poincar\'e supergroup, we conclude that the constraints \p{covchir} are also covariant.
Note that, with the constraints \p{covchir} taken into account, one may obtain from \p{nabladercomms} \cite{BG2}
\be
\big\{ \nabla_\alpha, \nabla_\beta   \big \}= \big \{ \bnabla_{\dot\alpha}, \bnabla_{\dot\beta}   \big \} =0,
\ee
and thus the subsequent action of the corresponding covariant derivatives on the constraints \p{covchir} will not produce any new ones.

The situation with the Bianchi identity (\ref{free}b) is more complicated. Indeed, after imposing the covariant chirality constraints \p{covchir}
the $d \theta^\alpha$ and $d\bar\theta{}_{\dot\alpha}$ parts
of the Cartan forms $\omega_S{}^\alpha$ and $\bar\omega_S{}_{\dot\alpha}$ \p{CF} contain only the quantities
\be\label{ad1}
\mV_{\alpha\beta}= \nabla_{(\alpha} \mpsi_{\beta)} \qquad \mbox{and} \qquad \mbV_{\dot\alpha\dot\beta}=- \bnabla_{(\dot\alpha} \mbpsi_{\dot\beta)}.
\ee
Of course, one may  impose the following covariant constraints
\be\label{eom}
\nabla_\alpha \mpsi^\alpha=0 \qquad \mbox{and} \qquad \bnabla{}^{\dot\alpha} \mbpsi{}_{\dot\alpha}=0.
\ee
But the problem is that these constraints put our theory on-shell \cite{BIK1}. Unfortunately, one cannot impose the constraint
\be\label{wrong}
\nabla_\alpha \mpsi^\alpha+ \bnabla{}^{\dot\alpha} \mbpsi{}_{\dot\alpha}=0,
\ee
because the first and second summands belong to the different Cartan forms and, therefore, being covariant can differ again on some covariant multipliers.
As the result, acting on \p{wrong} by $\nabla^2$ or $\bnabla{}^2$ we will get the equations of motion instead of identities \cite{BG2} and thus,
we come back to the equations \p{eom}. One may suppose that for the construction of the on-shell component action of D3-brane this is not a problem
and one may use the equations \p{eom}, which contain the Bianchi identity and the equations of motion. However, in such a situation we will be unable
to find a proper Bianchi identity on the bosonic components of $V_{(\alpha\beta)}, \bV_{(\dot\alpha\dot\beta)}$ and therefore it will be impossible
to find the invariant action. In the next Section we will demonstrate how this problem can be solved.
\setcounter{equation}{0}
\section{Curing the constraints}
The idea how to find the covariant Bianchi identity is rather simple. Let us suppose that we introduced into e tgame some additional bosonic superfield $\mphi$ which enters the new Cartan forms as
\be\label{bianchi1}
\Omega_S^\alpha \sim d\mpsi^\alpha - \im \mphi\, d\theta^\alpha + \ldots, \qquad
{\overline\Omega}_S{}_{\dot\alpha} \sim d \mbpsi_{\dot\alpha} +\im \mphi\, d \bar\theta_{\dot\alpha}+\ldots,
\ee
where $\ldots$ stands for all possible terms of higher orders in $\mphi$.
Note that such a modification is possible if some new generator $U$ commutes with the supersymmetry generators $Q$ and $S$  as follows
\be\label{nalg}
\big[ U, Q_\alpha \big]=S_\alpha, \quad  \big[ U, S_\alpha \big]=Q_\alpha, \quad \big[ U, \bQ_{\dot\alpha} \big]=- \bS_{\dot\alpha}, \quad
\big[ U, \bS_{\dot\alpha} \big]=- \bQ_{\dot\alpha}.
\ee
Therefore, the forms $\omega_Q$ and $\bar\omega_Q$ will be also changed, what is reflected in passing to a new set of covariant derivatives $\nabla \rightarrow \mD$.
Now, the covariant chirality constraints \p{covchir} will be replaced by new ones
\be\label{nchir}
\mbD_{\dot\alpha} \mpsi_\alpha =0, \quad \mD_\alpha \mbpsi_{\dot\alpha}=0,
\ee
while the constraints \p{eom} will be modified in the linear order of $\mphi$ as
\be\label{neom}
\mD_\alpha \mpsi^\alpha=\im \mphi, \quad \mbD{}^{\dot\alpha} \mbpsi{}_{\dot\alpha}=- \im \mphi.
\ee
Therefore, the covariant Bianchi identity will acquire the form
\be\label{cBianchi}
\mD_\alpha \mpsi^\alpha + \mbD{}^{\dot\alpha} \mbpsi{}_{\dot\alpha}=0,
\ee
while the equation
\be\label{neom2}
\mD_\alpha \mpsi^\alpha - \mbD{}^{\dot\alpha} \mbpsi{}_{\dot\alpha}=2 \im \mphi
\ee
will relate the first component of the superfield $\mphi$ with the auxiliary field of the $N=1, d=4$ off-shell vector supermultiplet. In other words, with the
Goldstone superfield $\mphi$ taken into account, we are in the off-shell situation and the newly defined Bianchi identity \p{cBianchi} are fully covariant.

In order to realize the idea which was sketched above, we extend the coset element $g$ \p{coset} as
\be\label{fcoset}
g\; \rightarrow \; g = e^{\im x^{\alpha\dot\alpha} P_{\alpha\dot\alpha}} e^{\im \left( \theta^\alpha Q_\alpha + \bar\theta_{\dot\alpha}\bQ^{\dot\alpha}   \right) }
e^{\im \left( \mpsi^\alpha S_\alpha + \mbpsi_{\dot\alpha}\bS^{\dot\alpha}   \right) } e^{\im \mphi U},
\ee
where $U$ is the generator defined in \p{nalg}, while $\mphi$ is the corresponding Goldstone superfield.
Such a type of modification of the coset element leads to new expressions of the Cartan forms. Firstly, the new Cartan forms corresponding to the generators
of unbroken supersymmetry $\Omega_Q$, ${\overline \Omega}{}_Q$ read
\be\label{nCFq}
\Omega_Q^\alpha =\cos\mphi\, d\theta^\alpha - \im \sin\mphi\, d \mpsi^\alpha, \quad
{\overline \Omega}_Q{}_{\dot\alpha} =\cos\mphi\, d \bar\theta_{\dot\alpha} + \im \sin\mphi\, d \mbpsi_{\dot\alpha}.
\ee
The modifications of these Cartan forms forced us to introduce the new covariant (with respect to $N=2, d=4$ Poincar\'e supergroup and $U$ transformations)
derivatives $\mD_\alpha , \mbD_{\dot\alpha}$ implicitly defined as
\bea
&& \nabla_\alpha = \cos\mphi\; \mD_\alpha - \im \sin\mphi\;\nabla_\alpha \mpsi^\beta \mD_\beta +
 \im \sin\mphi\; \nabla_\alpha \mbpsi_{\dot\beta} \mbD^{\dot\beta} , \nn \\
&& \bnabla{}^{\dot\alpha} = \cos\mphi\; \mbD{}^{\dot\alpha} -  \im \sin\mphi\; \bnabla{}^{\dot\alpha} \mpsi^\beta \mD_\beta +
 \im \sin\mphi\; \bnabla{}^{\dot\alpha} \mbpsi_{\dot\beta} \mbD^{\dot\beta} . \label{nCD}
\eea
Secondly, the forms $\Omega_S, {\overline \Omega}{}_S$ corresponding to the generators of broken supersymmetry  are also modified as
\bea\label{nCFs}
&&\Omega_S^\alpha =\frac{1}{\cos\mphi}\,\Big[  d\mpsi^\alpha - \im \sin\mphi\; \Omega_Q^\alpha \Big]=
\frac{1}{\cos\mphi}\,\Big[ \Omega_Q^\beta\Big( \mD_\beta \mpsi^\alpha - \im \delta_\beta^\alpha  \sin\mphi\Big )+
 {\overline \Omega}_Q{}_{\dot\beta} \mbD^{\dot\beta} \mpsi^\alpha +\omega_P^{\beta\dot\beta} \mD_{\beta\dot\beta} \mpsi^\alpha \Big], \nn \\
&&{\overline \Omega}_S{}_{\dot\alpha} =\frac{1}{\cos\mphi}\, \Big[ {\overline \Omega}_Q{}_{\dot\beta}\Big( \mbD{}^{\dot\beta} \mbpsi_{\dot\alpha}
+ \im \delta^{\dot\beta}_{\dot\alpha}  \sin\mphi\Big)+
\Omega_Q{}^{\beta} \mD_{\beta} \mbpsi_{\dot\alpha} +\omega_P^{\beta\dot\beta} \mD_{\beta\dot\beta} \mbpsi_{\dot\alpha} \Big].
\eea
Here, $\mD_{\alpha\dot\alpha}$ is a new modified space-time covariant derivative whose explicit form is unessential for what follows.

Now, one may impose the constraints on these forms. Nullifying the ${\overline \Omega}_Q$ projection of the form $\Omega_S$ and the $\Omega_Q$
projection of the form ${\overline \Omega}_S$ we will get the modified chirality constraints
\be\label{finchir}
\mbD_{\dot\alpha} \mpsi_\alpha =0, \qquad \mD_{\alpha} \mbpsi_{\dot\alpha}=0,
\ee
while nullifying the traces of the $\Omega_Q$   projection of the form $\Omega_S$ and the ${\overline \Omega}_Q$ projection of the form ${\overline \Omega}_S$ will result
in the constraints
\be\label{fineom}
\mD_\alpha \mpsi^\alpha = 2 \im \sin\mphi, \qquad \mbD{}^{\dot\alpha} \mbpsi_{\dot\alpha}=  -2\im \sin\mphi.
\ee
Thus, the constraints \p{finchir}, \p{fineom} have  the form we supposed   in \p{nchir}, \p{neom}.

Before going further, note that from the constrains \p{finchir} and, for example, from the first equation in \p{nCD} it follows that
\be\label{add1}
\nabla_\alpha \mbpsi_{\dot\beta} \big( \delta_{\dot\alpha}^{\dot\beta} -  \im \sin\mphi\; \mbD{}^{\dot\beta} \mbpsi_{\dot\alpha}\big) =0
\qquad \Rightarrow \qquad\nabla_\alpha \mbpsi_{\dot\beta}=0,
\ee
where the last equation is valid due to the non-degeneracy of the matrix in the parentheses in the first equation. Therefore, the new chirality conditions \p{finchir} are
equivalent to the previous ones \p{covchir}, and, again, they are in full agreement with the results of paper \cite{BG2}.

Taking into account  the constraints \p{finchir} and \p{add1}, one may represent the covariant derivatives $\mD_\alpha, \mbD_{\dot\alpha}$ in \p{nCD} as
\be\label{der1}
\mD_\alpha = \big({\cal A}^{-1}\big)_\alpha^\beta \; \nabla_\beta, \qquad \mbD{}^{\dot\alpha} =\big({\overline {\cal A}}{}^{-1}\big)^{\dot\alpha}_{\dot\beta} \,
\bnabla{}^{\dot\beta},
\ee
where
\be\label{A}
\big(\cal A\,\big)_\alpha^\beta = \cos\mphi\; \delta_\alpha^\beta - \im \sin\mphi\;\nabla_\alpha \mpsi^\beta, \quad
\big(\overline{\cal A}\,\big)_{\dot\alpha}^{\dot\beta} = \cos\mphi\; \delta_{\dot\alpha}^{\dot\beta} + \im \sin\mphi\;\bnabla{}^{\dot\beta} \mbpsi_{\dot\alpha}.
\ee
Thus, we conclude that the proper covariant Bianchi identity reads
\be\label{BF}
\big({\cal A}^{-1}\big)_\alpha^\beta \; \nabla_\beta \mpsi^\alpha + \big({\overline {\cal A}}{}^{-1}\big)^{\dot\alpha}_{\dot\beta} \; \bnabla{}^{\dot\beta} \mbpsi_{\dot\alpha} =0 ,
\ee
where the matrices $\cal A$, $\overline{\cal A}$ are defined in \p{A}, while the equation defining the superfield $\mphi$ has the form
\be\label{u}
\big({\cal A}^{-1}\big)_\alpha^\beta \; \nabla_\beta \mpsi^\alpha - \big({\overline {\cal A}}{}^{-1}\big)^{\dot\alpha}_{\dot\beta} \;
\bnabla{}^{\dot\beta} \mbpsi_{\dot\alpha} =4 \im \sin\mphi.
\ee

In practice, the equations \p{BF} are not convenient, because they  define the Bianchi identity in an implicit way.
In order to get a more convenient form of the Bianchi identity, one has to substitute, for example, into the first equation of \p{fineom}
\be\label{test1}
\mD_\alpha \mpsi^\alpha =\big( \cA^{-1}\big)_\alpha^\beta \nabla_\beta\mpsi^\alpha = 2 \im \sin\mphi,
\ee
the expression for $\nabla_\beta\mpsi^\alpha$ which follows from \p{A}
\be\label{test2}
\nabla_\beta\mpsi^\alpha =-\frac{\im}{\sin\mphi}\big(\cos\mphi\;\delta_\beta^\alpha-\big( \cA\big)^\alpha_\beta\big)\;\;\; \Rightarrow\;\;\;
\mD_\alpha \mpsi^\alpha = -\frac{\im}{\sin\mphi}\big( \cos\mphi\; \mbox{Tr}\big(\cA^{-1}\big) -2 \big)= 2 \im \sin\mphi \,.
\ee
The solution of the equation \p{test2} may be easily found with respect to "$\tan \mphi$" and it reads
\be\label{test3}
\tan\mphi =-\im\, \frac{B}{1+\sqrt{1-B^2}}\,, \qquad B\equiv \frac{\nabla_\alpha\mpsi^\alpha}{1+\frac{1}{2} \nabla^\beta\mpsi^\gamma\; \nabla_\beta \mpsi_\gamma}\,.
\ee
Repeating the same calculations for $\bnabla^{\dot\alpha} \mbpsi_{\dot\beta}$, we will get the conjugated expression
\be\label{test4}
\tan\mphi =\im\, \frac{\bB}{1+\sqrt{1-\bB^2}}\,, \qquad \bB\equiv \frac{\bnabla^{\dot\alpha}\mbpsi_{\dot\alpha}}{1+\frac{1}{2}
\bnabla^{\dot\beta}\mbpsi^{\dot\gamma}\; \bnabla_{\dot\beta} \mbpsi_{\dot\gamma}}\,,
\ee
and, therefore
\be\label{test5}
\frac{B}{1+\sqrt{1-B^2}}+\frac{\bB}{1+\sqrt{1-\bB^2}}=0\qquad \Rightarrow \qquad B+\bB=0.
\ee
Thus, the suitable covariant version of the Bianchi identity reads
\be\label{BI}
 \frac{\nabla_\alpha\mpsi^\alpha}{1+\frac{1}{2} \nabla^\beta\mpsi^\gamma\; \nabla_\beta \mpsi_\gamma} + \frac{\bnabla^{\dot\alpha}\mbpsi_{\dot\alpha}}{1+\frac{1}{2}
\bnabla^{\dot\beta}\mbpsi^{\dot\gamma}\; \bnabla_{\dot\beta} \mbpsi_{\dot\gamma}}=0.
\ee

It is worth to compare the Bianchi identity \p{BI} with those which have been found in the expansion on fields up to the third order in \cite{BG2}.
Expanding the Bianchi identity \p{BI} in a series of the fields we will get
\be\label{test6}
\nabla_\alpha \mpsi^\alpha -\frac{1}{2}\, \nabla_\alpha \mpsi^\alpha \nabla^\beta\mpsi^\gamma\; \nabla_\gamma \mpsi_\beta -\frac{1}{2}\, \big(\nabla_\alpha \mpsi^\alpha\big)^3+
\bnabla^{\dot\alpha}\mbpsi_{\dot\alpha} -\frac{1}{2}\, \bnabla^{\dot\alpha}\mbpsi_{\dot\alpha}\bnabla^{\dot\beta}\mbpsi^{\dot\gamma}\;
\bnabla_{\dot\gamma} \mbpsi_{\dot\beta}    -\frac{1}{2}\,\big( \bnabla^{\dot\alpha}\mbpsi_{\dot\alpha}\big)^3 = {\cal O}(\mpsi^5).
\ee
Keeping in mind that on the surface of the constraints \p{test6}
\be\label{test7}
-\frac{1}{2}\, \big(\nabla_\alpha \mpsi^\alpha\big)^3 -\frac{1}{2}\,\big( \bnabla^{\dot\alpha}\mbpsi_{\dot\alpha}\big)^3 = {\cal O}\big(\mpsi^5 \big),
\ee
we will get to the third order on the fields
\be\label{test8}
\nabla_\alpha \mpsi^\alpha -\frac{1}{2}\, \nabla_\alpha \mpsi^\alpha \nabla^\beta\mpsi^\gamma\; \nabla_\gamma \mpsi_\beta+
\bnabla^{\dot\alpha}\mbpsi_{\dot\alpha} -\frac{1}{2}\, \bnabla^{\dot\alpha}\mbpsi_{\dot\alpha}\bnabla^{\dot\beta}\mbpsi^{\dot\gamma}\;
\bnabla_{\dot\gamma} \mbpsi_{\dot\beta} = {\cal O} \big(\mpsi^5 \big),
\ee
which exactly matches the Bianchi identity obtained in \cite{BG2}.

The last step is to check that the action of $\nabla^2$ and $\bnabla{}^2$ of the \p{BI} leads to the identities.
To prove this, it is found to be convenient to represent the results
of the action on $\nabla_\alpha$ and $\bnabla_{\dot\alpha}$ on \p{BI} as
\bea\label{nabla2}
&&\nabla^2 \mpsi_\alpha = -4 \im\, \frac{1+\frac{1}{2} \nabla^\mu \mpsi^\gamma \nabla_\mu \mpsi_\gamma}{1+\frac{1}{2} \bnabla^{\dot\mu} \mbpsi^{\dot\gamma}
\bnabla_{\dot\mu} \mbpsi_{\dot\gamma}} \left[ \nabla_{\alpha\dot\alpha} \mbpsi{}^{\dot\alpha}
-\bnabla^{\dot\beta}\mbpsi^{\dot\alpha} \nabla_{\beta} \mpsi_{\alpha}\nabla^{\beta}{}_{\dot\beta} \mbpsi_{\dot\alpha} \right], \nn\\
&&\bnabla^2 \mbpsi_{\dot\alpha} = 4 \im\, \frac{1+\frac{1}{2} \bnabla^{\dot\mu} \mbpsi^{\dot\gamma} \bnabla_{\dot\mu} \mbpsi_{\dot\gamma}}
{1+\frac{1}{2} \nabla^\mu \mpsi^\gamma \nabla_\mu \mpsi_\gamma} \left[ \nabla_{\alpha\dot\alpha} \mpsi{}^{\alpha}
-\nabla^{\beta}\mpsi^{\alpha} \bnabla_{\dot\beta} \mbpsi_{\dot\alpha}\nabla_{\beta}{}^{\dot\beta} \mpsi_{\alpha} \right].
\eea
Now, it is a matter of straightforward but still quite lengthly calculations to check that the r.h.s. of the equations \p{nabla2} are covariantly chiral
and antichiral, respectively. Thus we see that the Bianchi identities \p{BI} are correct: they are covariant with respect to all symmetries
and do not imply the equations of motion.
\setcounter{equation}{0}
\section{On-shell component action}
Having at hands the covariant Bianchi identity \p{BI}, one may try to construct the component or even the superfield action. The resulting action is expected to be rather
complicated, similarly to the superfield actions constructed within linear realization approaches in \cite{BG2, RT}. Instead, here we are going to demonstrate
that the on-shell component action of D3-brane has a very simple form. To construct such an action one has, firstly, to find the expressions for the auxiliary
components in the superfields $\mpsi^\alpha$, $\mbpsi_{\dot\alpha}$ and the  Bianchi identity on the bosonic components $V_{\alpha\beta}$,
$\bV_{\dot\alpha\dot\beta}$ \p{ad1} which follow from \p{BI}. Then one has to write the most general component action, invariant with respect
to broken supersymmetry and, finally, to fix the Lagrangian by imposing the invariance with respect to unbroken supersymmetry.
\subsection{On-shell content and the bosonic Bianchi identity}
It follows from the previous analysis, that going on-shell means nullifying the superfield $\mphi$:
\be\label{onshell1}
\mphi = 0 \quad \Rightarrow \quad \nabla_\alpha \mpsi^\alpha = \bnabla^{\dot\alpha} \mbpsi_{\dot\alpha} =0.
\ee
Thus, the on-shell vector supermultiplet contains the following physical components
\be\label{onshell2}
\psi_\alpha = \mpsi_\alpha|, \quad \bpsi_{\dot\alpha} = \mbpsi_{\dot\alpha}|, \quad V_{\alpha\beta} = \nabla_{(\alpha} \mpsi_{\beta)}|, \quad
\bV_{\dot\alpha\dot\beta} = -\bnabla_{(\dot\alpha} \mbpsi_{\dot\beta)}|,
\ee
where, as usual, $|$ means $\theta=\bar\theta=0$ limit. The definitions \p{onshell2} have to be supplied by the Bianchi identity needed to treat
the fields $V_{\alpha\beta}, \bV_{\dot\alpha\dot\beta}$ as the components of the field strength. Before the construction of the Bianchi identity in a full
generality, it is instructive to derive their bosonic limit. It can be done by acting with the derivatives $\nabla_\alpha$ and $\bnabla_{\dot\alpha}$ on \p{BI} and
taking the limit $\psi, \bpsi \rightarrow 0$ after calculations. Doing so, we will get
\bea\label{BIbos1}
\frac{\partial_{\beta\dot\alpha} V_\alpha^\beta + V_\beta^\gamma \, \bV_{\dot\alpha}^{\dot\gamma} \, \partial_{\gamma\dot\gamma}V_\alpha^\beta}{1 + \frac{1}{2}V^2 } - \frac{\partial_{\alpha\dot\beta}\bV^{\dot\beta}_{\dot\alpha} + V_\alpha^\gamma \,
\bV_{\dot\beta}^{\dot\gamma} \, \partial_{\gamma\dot\gamma} \bV^{\dot\beta}_{\dot\alpha}}{1 + \frac{1}{2}\bV^2 } \equiv \big({\tt BId}  \big)_{\alpha\dot\alpha} =0.
\eea
It has been already clarified in \cite{BIK1}, that these conditions can be transformed in the conventional form of the Bianchi identity.
Indeed, evaluating the combination $\big ({\tt BId} \big)_{\alpha\dot\alpha} + V_\alpha^\beta \, \bV_{\dot\alpha}^{\dot\beta}\, \big ({\tt BId} \big)_{\beta\dot\beta}$
and multiplying it by function
$$
\frac{\left( 1+ \frac{1}{2} V^2 \right)\left( 1 + \frac{1}{2} \bV^2 \right)}{\left( 1 -\frac{1}{4} V^2 \bV^2  \right)^2},
$$
we will get
\be\label{BIbos2}
\partial_{\beta\dot\alpha}\left[ V_\alpha^\beta \frac{1 + \frac{1}{2}\bV^2}{1 - \frac{1}{4}V^2 \bV^2}    \right]
- \partial_{\alpha\dot\beta}\left[ \bV_{\dot\alpha}^{\dot\beta} \frac{1 + \frac{1}{2}V^2}{1 - \frac{1}{4}V^2 \bV^2}    \right] =0.
\ee
Thus, the bosonic fields $V_{\alpha\beta}$ and $\bV_{\dot\alpha\dot\beta}$ are indeed the components of the field strength, as it should be.
\subsection{Fixing the action}
Within our approach, the construction of the invariant component on-shell action can be performed in two steps:
\begin{itemize}
\item Firstly, one has to construct the most general action invariant with respect to broken supersymmetry
\item Secondly, the action has to be fixed by imposing its invariance with respect to unbroken supersymmetry.
\end{itemize}
To perform the first step, let us introduce the $\theta=0$ limit of the vierbein $E_{\alpha\dot\alpha}^{\beta\dot\beta}$ and the space-time
derivative $\nabla_{\alpha\dot\alpha}$ defined in \p{CD}
\be\label{CDonshell}
\cD_{\alpha\dot\alpha} = \nabla_{\alpha\dot\alpha}|_{\theta=0} = \left( \cE^{-1}\right)_{\alpha\dot\alpha}^{\beta\dot\beta} \; \partial_{\beta\dot\beta}, \quad
\cE_{\alpha\dot\alpha}^{\beta\dot\beta}=E_{\alpha\dot\alpha}^{\beta\dot\beta}|_{\theta=0} =\delta_\alpha^\beta \delta_{\dot\alpha}^{\dot\beta}
-\im \psi{}^\beta \partial_{\alpha\dot\alpha} \bpsi{}^{\dot\beta}-
\im \bpsi{}^{\dot\beta} \partial_{\alpha\dot\alpha} \psi{}^{\beta}.
\ee
Clearly, the vierbein $\cE_{\alpha\dot\alpha}^{\beta\dot\beta}$ and the space-time derivative $\cD_{\alpha\dot\alpha}$ are covariant with respect to broken supersymmetry. Now, keeping in  mind that, with respect to broken $(S)$ supersymmetry, the coordinates and the Goldstone superfields transform as
\be\label{Str}
\delta_S x^{\alpha\dot\alpha} = \im \left( \varepsilon^\alpha \mbpsi^{\dot\alpha} + \bar\varepsilon{}^{\dot\alpha} \mpsi^\alpha\right), \quad
\delta_S \theta_\alpha =\delta_S \bar\theta_{\dot\alpha}=0, \quad \delta_S \mpsi_\alpha= \varepsilon_\alpha, \quad \delta_S \mbpsi_{\dot\alpha} = \bar\varepsilon_{\dot\alpha},
\ee
one may immediately conclude that the most general component action which is invariant under transformations \p{Str} reads
\be\label{act1}
S= \int d^4 x \det \cE \; G( V^2, \bV^2 ).
\ee
Here, the function $G$ is an arbitrary function whose explicit form has to be fixed by the invariance of the action \p{act1} with respect to unbroken supersymmetry.

In order to fix the function $G$ in the action \p{act1} one has to know the transformation properties of the components \p{onshell2} under unbroken supersymmetry, which
are defined in a standard way as
\bea\label{Qtrans1}
&&
\delta_Q f = -\big(\epsilon^\alpha D_\alpha + \bar\epsilon_{\dot\alpha}\bD^{\dot\alpha}\big) \mf |_{\theta \rightarrow 0}
= - \big(\epsilon^\alpha \nabla_\alpha + \bar\epsilon_{\dot\alpha}\bnabla^{\dot\alpha}\big)\mf |_{\theta \rightarrow 0}
+ H^{\lambda\dot\lambda}\partial_{\lambda\dot\lambda}f, \nn \\
&&
H^{\lambda\dot\lambda} = -\im\, \big( \epsilon^{\gamma} \bpsi^{\dot\lambda} V_\gamma^\lambda -\bar\epsilon_{\dot\gamma}\psi^\lambda \bV^{\dot\gamma\dot\lambda}\big).
\eea
Thus, we have (we write explicitly only the $\epsilon^\alpha$-part)
\bea\label{Qsusypsi}
&&\delta_Q \psi^\alpha = -\epsilon^\gamma V_\gamma^\alpha + H^{\lambda\dot\lambda} \partial_{\lambda\dot\lambda} \psi^\alpha, \quad
\delta_Q \bpsi_{\dot\alpha} =  H^{\lambda\dot\lambda} \partial_{\lambda\dot\lambda} \bpsi_{\dot\alpha}, \nn \\
&&\delta_Q \det\cE =\partial_{{\lambda\dot\lambda}} \big( H^{\lambda\dot\lambda} \det\cE  \big)
+ 2\im \, \epsilon^\gamma \cD_{\alpha\dot\alpha} \bpsi^{\dot\alpha} V_{\gamma}^{\alpha}\; \det\cE ,\nn
\eea
and
\bea\label{QsusyV}
&&
\delta_Q V_{\alpha\beta} = 2\im\, \frac{1+ \frac{1}{2}V^2}{1+ \frac{1}{2}\bV^2}\,\epsilon_\beta\, \big( \cD_{\alpha\dot\alpha} \bpsi^{\dot\alpha}
+ V^{\lambda}_\alpha \bV_{\dot\alpha}^{\dot\lambda} \cD_{\lambda\dot\lambda} \bpsi^{\dot\alpha}   \big)
+ H^{\lambda\dot\lambda}\partial_{\lambda\dot\lambda} V_{\alpha\beta}, \nn \\
&&
\delta_Q \bV_{\dot\alpha\dot\beta} = 2\im\, \epsilon^\gamma \big( \cD_{\gamma\dot\beta}\bpsi_{\dot\alpha}
+ V_\gamma^\lambda \bV_{\dot\beta}^{\dot\lambda}\cD_{\lambda\dot\lambda}\bpsi_{\dot\alpha}\big) +
H^{\lambda\dot\lambda}\partial_{\lambda\dot\lambda} \bV_{\dot\alpha\dot\beta}, \nn \\
&&
\delta_Q V^2 =  -4\im\, \frac{1+ \frac{1}{2}V^2}{1+ \frac{1}{2}\bV^2}\, \epsilon^\beta\,
\big( V^\alpha_{\beta} \cD_{\alpha\dot\alpha} \bpsi^{\dot\alpha} - \frac{1}{2}\,V^2\, \bV_{\dot\alpha}^{\dot\lambda}\cD_{\beta\dot\lambda}\bpsi^{\dot\alpha} \big)+
H^{\lambda\dot\lambda}\partial_{\lambda\dot\lambda}V^2, \\
&&
\delta_Q \bV^2 =  - 4\im\, \epsilon^\gamma\, \big( \bV_{\dot\alpha}^{\dot\beta} \cD_{\gamma\dot\beta} \bpsi^{\dot\alpha}
- \frac{1}{2}\,\bV^2  V_\gamma^\lambda \cD_{\lambda\dot\alpha}\bpsi^{\dot\alpha} \big)+H^{\lambda\dot\lambda}\partial_{\lambda\dot\lambda}\bV^2.\nn
\eea
Therefore, the variation of Lagrangian entering in the ansatz \p{act1} reads
\bea\label{Qsusylagr}
&&
\delta_Q {\cal L} = \delta_Q \big[ \det\cE G\big(V^2, \bV^2 \big)\big] \\
&&
= -\im \det\cE \epsilon^\beta \cD_{\alpha\dot\alpha}\bpsi^{\dot\alpha} V_{\beta}^{\alpha} \bigg[  2 G
-4\, \frac{1+\frac{1}{2}V^2 }{1 + \frac{1}{2}\bV^2} \frac{\partial{G}}{\partial{V^2}} + 2 \bV^2 \frac{\partial{G}}{\partial{\bV^2}}   \bigg] +
+ \im \det\cE \epsilon^\beta \cD_{\beta\dot\beta}\bpsi^{\dot\alpha}\bV_{\dot\alpha}^{\dot\beta} \bigg[ 4\,\frac{\partial{G}}{\partial{\bV^2}}
- 2V^2 \frac{\partial{G}}{\partial{V^2}} \frac{1+ \frac{1}{2}V^2 }{1 + \frac{1}{2}\bV^2}  \bigg]. \nn
\eea
While obtaining \p{Qsusylagr} we neglected the $H^{\alpha\dot\alpha}$-dependent terms which combine into a total divergence.

It is funny that to find the function $G$ it is enough to consider the variation \p{Qsusylagr} in the first order in the fermions $\psi, \bpsi$, i.e.
\be\label{Qvar2}
\delta_Q {\cal L} = -\im\, \epsilon^\beta \partial_{\alpha\dot\alpha}\bpsi^{\dot\alpha} V_{\beta}^{\alpha}
\bigg[  2 G -4\, \frac{1+\frac{1}{2}V^2 }{1 + \frac{1}{2}\bV^2} \frac{\partial{G}}{\partial{V^2}} + 2 \bV^2 \frac{\partial{G}}{\partial{\bV^2}}   \bigg]
+ \im\, \epsilon^\beta \partial_{\beta\dot\beta}\bpsi^{\dot\alpha}\bV_{\dot\alpha}^{\dot\beta}
\bigg[ 4\,\frac{\partial{G}}{\partial{\bV^2}} - 2V^2 \frac{\partial{G}}{\partial{V^2}} \frac{1+ \frac{1}{2}V^2 }{1 + \frac{1}{2}\bV^2}  \bigg].
\ee
Integrating by parts in \p{Qvar2} one may represent the variation of the Lagrangian as
\be\label{Qvar3}
\delta_Q {\cal L} = \im\, \epsilon^\beta \bpsi^{\dot\alpha}\bigg\{ \partial_{\alpha\dot\alpha}\bigg[  V_{\beta}^{\alpha}
\bigg(  2 G -4\, \frac{1+\frac{1}{2}V^2 }{1 + \frac{1}{2}\bV^2} \frac{\partial{G}}{\partial{V^2}} + 2 \bV^2 \frac{\partial{G}}{\partial{\bV^2}}   \bigg)\bigg]
 - \partial_{\beta\dot\beta}\bigg[ \bV_{\dot\alpha}^{\dot\beta} \bigg( 4\,\frac{\partial{G}}{\partial{\bV^2}}
 - 2V^2 \frac{\partial{G}}{\partial{V^2}} \frac{1+ \frac{1}{2}V^2 }{1 + \frac{1}{2}\bV^2}  \bigg)\bigg]\bigg\}.
\ee
Now, it is easy to check that if we will choose the function $G$ as
\be\label{G}
G = 1 + \frac{\big( 1 + \frac{1}{2}V^2   \big) \big( 1+ \frac{1}{2}\bV^2   \big)}{1 - \frac{1}{4}V^2\bV^2 },
\ee
the expression in the curly brackets in \p{Qvar3} will coincide with the l.h.s. of the bosonic Bianchi identity \p{BIbos2}, and therefore, to this order, the
action
\be\label{act2}
S= \int d^4 x \det \cE \bigg[  1 + \frac{\big( 1 + \frac{1}{2}V^2   \big) \big( 1+ \frac{1}{2}\bV^2   \big)}{1 - \frac{1}{4}V^2\bV^2 }\bigg]
\ee
will be invariant with respect to both broken and unbroken supersymmetries.

Note that the action \p{act2} is completely fixed. So, if everything was correct this action has to be invariant under unbroken supersymmetry transformations
\p{Qsusylagr} with all fermions taken into account. To prove this one needs, firstly, to derive the complete Bianchi identity.
\subsection{Complete Bianchi identity}
Repeating the same steps as in deriving the bosonic Bianchi identity \p{BIbos2} one may get
\bea\label{BI2}
&& \cD_{\beta\dot\alpha} \bigg[ V_\alpha^\beta \frac{1 + \frac{1}{2}\bV^2}{1 - \frac{1}{4}V^2 \bV^2}     \bigg]
-  \cD_{\alpha\dot\beta} \bigg[ \bV_{\dot\alpha}^{\dot\beta} \frac{1 + \frac{1}{2}V^2}{1 - \frac{1}{4}V^2 \bV^2}    \bigg] + \nn \\
&&+2 \im\, \frac{1 + \frac{1}{2}\bV^2}{1 - \frac{1}{4}V^2 \bV^2} \Big[ V_\alpha^\beta \cD_{\lambda\dot\lambda} \psi^{\lambda} \cD_{\beta\dot\alpha}\bpsi^{\dot\lambda}
- V_\alpha^\beta \cD_{\beta\dot\alpha} \psi^{\lambda} \cD_{\lambda\dot\lambda} \bpsi^{\dot\lambda}   - V^{\sigma\rho} \cD_{\sigma\dot\gamma}
\psi_{\alpha} \cD_{\rho}^{\dot\gamma} \bpsi_{\dot\alpha}  \Big]- \\
&&- 2\im\,  \frac{1 + \frac{1}{2}V^2}{1 - \frac{1}{4}V^2 \bV^2} \Big[ \bV_{\dot\alpha}^{\dot\beta} \cD_{\lambda\dot\lambda}\psi^{\lambda} \cD_{\alpha\dot\beta}\bpsi^{\dot\lambda}
- \bV_{\dot\alpha}^{\dot\beta} \cD_{\alpha\dot\beta}\psi^{\lambda} \cD_{\lambda\dot\lambda} \bpsi^{\dot\lambda}
- \bV^{\dot\sigma\dot\rho} \cD_{\gamma \dot\rho}\psi_\alpha \cD_{\dot\sigma}^\gamma\bpsi_{\dot\alpha} \Big] =0.\nn
\eea
Then, multiplying this equation by $\det \cE$, one finds that it can be rewritten as
\bea\label{BI3}
\partial_{\gamma\dot\gamma}\bigg[ \det \cE \left( \cE^{-1}   \right)^{\gamma\dot\gamma}_{\beta\dot\alpha} V_{\alpha}^\beta \frac{1 + \frac{1}{2}\bV^2}{1 - \frac{1}{4}V^2 \bV^2}
-  \det \cE \left( \cE^{-1}   \right)^{\gamma\dot\gamma}_{\alpha\dot\beta}     \bV_{\dot\alpha}^{\dot\beta}\frac{1 + \frac{1}{2}V^2}{1 - \frac{1}{4}V^2 \bV^2} \bigg] = \nn \\
= 2\im \det \cE \bigg[ \frac{1 + \frac{1}{2}\bV^2}{1 - \frac{1}{4}V^2 \bV^2} V^{\sigma\rho} \cD_{\sigma\dot\gamma}
\psi_{\alpha} \cD_{\rho}^{\dot\gamma} \bpsi_{\dot\alpha} -  \frac{1 + \frac{1}{2}V^2}{1
- \frac{1}{4}V^2 \bV^2} \bV^{\dot\sigma\dot\rho} \cD_{\gamma \dot\rho}\psi_\alpha \cD_{\dot\sigma}^\gamma\bpsi_{\dot\alpha}   \bigg].
\eea
This representation of the Bianchi identity is quite convenient but it does not have the standard form  for the genuine field strength
\be\label{BI4}
\partial_{\beta\dot\alpha} F^\beta_\alpha - \partial_{\alpha\dot\beta}\bF^{\dot\beta}_{\dot\alpha} = 0.
\ee
To bring the Bianchi identity \p{BI3} to the standard form, firstly, one has to rewrite the second line in \p{BI3} as
\be\label{BI5}
2 \im \partial_{\gamma\dot\gamma}\bigg[ \det\cE\, \frac{1 + \frac{1}{2}\bV^2}{1 - \frac{1}{4}V^2 \bV^2} V^{\rho\sigma} \big(  \cE^{-1}\big)^{\gamma\dot\gamma}_{\rho\dot\mu}
\big( \psi_\alpha \cD_\sigma^{\dot\mu} \bpsi_{\dot\alpha} + \bpsi_{\dot\alpha} \cD_{\sigma}^{\dot\mu} \psi_\alpha    \big)
-  \det\cE \frac{1 + \frac{1}{2}V^2}{1 - \frac{1}{4}V^2 \bV^2} \bV^{\dot\rho\dot\sigma} \big(\cE^{-1}\big)^{\gamma\dot\gamma}_{\mu\dot\rho}
\big( \psi_\alpha \cD_{\dot\sigma}^{\mu} \bpsi_{\dot\alpha} + \bpsi_{\dot\alpha} \cD_{\dot\sigma}^{\mu} \psi_\alpha    \big) \bigg].
\ee
It is not obvious that this expression coincides with the second line in \p{BI3}. The differences of these two expressions contains undifferentiated
fermions $\psi$ and $\bpsi$. Let us consider the $\psi$-dependent terms (the consideration of $\bpsi$-dependent terms going in the same way) which read
\bea\label{BI6}
&& 2 \im\, \psi_\alpha \partial_{\gamma\dot\gamma}\bigg[ \det\cE \big( \cE^{-1}  \big)^{\gamma\dot\gamma}_{\rho\dot\mu}\, V^{\rho}_{\sigma}
\frac{1 + \frac{1}{2}\bV^2}{1 - \frac{1}{4}V^2 \bV^2}  - \det\cE \big( \cE     ^{-1}\big)^{\gamma\dot\gamma}_{\sigma\dot\rho}\, \bV^{\dot\rho}_{\dot\mu}
\frac{1 + \frac{1}{2}V^2}{1 - \frac{1}{4}V^2 \bV^2}       \bigg]\cD^{\sigma\dot\mu}\bpsi_{\dot\alpha}+ \nn \\
&&+2 \im\,  \psi_\alpha \det\cE  \bigg[ \frac{1 + \frac{1}{2}\bV^2}{1 - \frac{1}{4}V^2 \bV^2}\, V^{\rho\sigma} \cD_{\rho\dot\mu}\cD^{\dot\mu}_\sigma \bpsi_{\dot\alpha}
- \frac{1 + \frac{1}{2}V^2}{1 - \frac{1}{4}V^2 \bV^2}\,\bV^{\dot\rho\dot\sigma} \cD_{\mu\dot\rho}\cD^\mu_{\dot\sigma} \bpsi_{\dot\alpha} \bigg].
\eea
Now, one should use the identity \p{BI3} again to rewrite the first line in \p{BI6}, and note, that the symmetrized double derivatives in the second line are
actually commutators, $\cD_{(\sigma\dot\gamma} \cD^{\dot\gamma}_{\rho)} = \frac{1}{2} \big[ \cD_{(\sigma\dot\gamma}, \cD^{\dot\gamma}_{\rho)}   \big]$,
that are already known \p{nabladercomms}. Taking all this into account, one may check that all terms in \p{BI6} completely cancel each other.

Before combining the expressions \p{BI3} and \p{BI5} together, note that combinations of fermions and their derivatives in \p{BI5} can be written in terms of
the inverse fierbeins $\big( \cE^{-1}   \big)^{\beta\dot\beta}_{\alpha\dot\alpha}$ as
\be\label{BI7}
\im\, \big( \psi_\alpha \cD_\sigma^{\dot\mu} \bpsi_{\dot\alpha} + \bpsi_{\dot\alpha} \cD_{\sigma}^{\dot\mu} \psi_\alpha    \big)
= \epsilon_{\alpha\beta}\epsilon_{\dot\alpha\dot\beta}\epsilon^{\dot\mu\dot\sigma} \Big[ \big( \cE^{-1}   \big)^{\beta\dot\beta}_{\sigma\dot\sigma}
- \delta^\beta_\sigma \delta^{\dot\beta}_{\dot\sigma}   \Big].
\ee
Finally, the Bianchi identity acquires the form
\bea\label{BI8}
 \partial_{\gamma\dot\gamma}\bigg[ \det\cE\,  V^{\rho\sigma}  \frac{1 + \frac{1}{2}\bV^2}{1 - \frac{1}{4}V^2 \bV^2}\,
 \epsilon^{\dot\mu\dot\sigma} \big( \cE^{-1}   \big)^{\gamma\dot\gamma}_{\rho\dot\mu}\big(\cE^{-1}\big)^{\beta\dot\beta}_{\sigma\dot\sigma}
 - \det\cE\, \bV^{\dot\rho\dot\sigma}   \frac{1 + \frac{1}{2}\bV^2}{1 - \frac{1}{4}V^2 \bV^2}\,
 \epsilon^{\mu\sigma}\big( \cE^{-1}   \big)^{\gamma\dot\gamma}_{\mu\dot\rho}\big(\cE^{-1}    \big)^{\beta\dot\beta}_{\sigma\dot\sigma}  \bigg]=0.
\eea
The close inspection of terms
\be
V^{\rho\sigma} \epsilon^{\dot\mu\dot\sigma} \big( \cE^{-1}   \big)^{\gamma\dot\gamma}_{\rho\dot\mu}\big(\cE^{-1}\big)^{\beta\dot\beta}_{\sigma\dot\sigma} \quad \mbox{and}\quad
\bV^{\dot\rho\dot\sigma} \epsilon^{\mu\sigma}\big( \cE^{-1}   \big)^{\gamma\dot\gamma}_{\mu\dot\rho}\big(\cE^{-1}    \big)^{\beta\dot\beta}_{\sigma\dot\sigma}
\ee
shows that they are antisymmetric over interchanging of  the indices $\{ \beta\dot\beta \}$ and $\{\gamma\dot\gamma  \}$.
Therefore, the product of two $\cE$ terms splits into two parts which are proportional to $\epsilon^{\beta\gamma}$ and
to $\epsilon^{\dot\beta\dot\gamma}$, respectively.  Taking this into account, one may finally write down the genuine physical field strengths
\bea\label{Fphys}
F^{\alpha\beta} &=& -\frac{1}{2}\bigg\{ \det\cE \frac{1 + \frac{1}{2}\bV^2}{1 - \frac{1}{4}V^2 \bV^2} V^{\rho\sigma}
\Big[ \big( \cE^{-1}  \big)^{\beta\dot\lambda}_{\rho\dot\mu}\big( \cE^{-1}  \big)^{\alpha\dot\mu}_{\sigma\dot\lambda}
- \big( \cE^{-1}  \big)^{\beta\dot\lambda}_{\rho\dot\lambda}\big(\cE^{-1}  \big)^{\alpha\dot\mu}_{\sigma\dot\mu}    \Big] -  \nn \\
&&- \det\cE \frac{1 + \frac{1}{2}V^2}{1 - \frac{1}{4}V^2 \bV^2} \bV^{\dot\rho\dot\sigma} \Big [\epsilon_{\dot\lambda\dot\tau}\epsilon^{\mu\nu}
\big( \cE^{-1}  \big)^{\beta\dot\lambda}_{\mu\dot\rho}\big( \cE^{-1}  \big)^{\alpha\dot\tau}_{\nu\dot\sigma}\Big] \bigg\}, \nn \\
\bF^{\dot\alpha\dot\beta} &=& -\frac{1}{2} \bigg \{ \det\cE \frac{1 + \frac{1}{2}V^2}{1 - \frac{1}{4}V^2 \bV^2}\bV^{\dot\rho\dot\sigma}
\Big[ \big( \cE^{-1}  \big)^{\lambda\dot\beta}_{\mu\dot\rho}\big( \cE^{-1}  \big)^{\mu\dot\alpha}_{\lambda\dot\sigma}
- \big( \cE^{-1}  \big)^{\lambda\dot\beta}_{\lambda\dot\rho}\big(\cE^{-1}  \big)^{\mu\dot\alpha}_{\mu\dot\sigma}    \Big] - \nn \\
&&- \det\cE \frac{1 + \frac{1}{2}\bV^2}{1 - \frac{1}{4}V^2 \bV^2} V^{\rho\sigma} \Big [\epsilon_{\lambda\tau}\epsilon^{\dot\mu\dot\nu}
\big( \cE^{-1} \big)^{\lambda\dot\beta}_{\rho\dot\mu} \big( \cE^{-1}  \big)^{\tau\dot\alpha}_{\sigma\dot\nu}\Big] \bigg\},
\eea
which obey the conventional Bianchi identity \p{BI4}.

Let us note that the contribution of $\bV^{\dot\rho\dot\sigma}$ in $F^{\alpha\beta}$ is purely fermionic; indeed, neglecting all fermions,
the respective combination of $\cE^{-1}$-symbols reduces to $\epsilon_{\dot\rho\dot\sigma}$ and thus disappears.
However, the terms quadratic in fermions lead to a non-zero contribution:
\bea\label{Equadr}
\im \bV^{\dot\rho\dot\sigma}\Big[\psi^\beta \cD^\alpha_{\dot\rho}\bpsi_{\dot\sigma} + \bpsi_{\dot\sigma}\cD^\alpha_{\dot\rho}\psi^\beta
+ \psi^{\alpha} \cD^\beta_{\dot\sigma}\bpsi_{\dot\rho} + \bpsi_{\dot\rho} \cD^{\beta}_{\dot\sigma}\psi^\alpha \Big].
\eea

Let us finally note, that the slightly complicated form of the genuine field strength in \p{Fphys} is a result of using the spinor notation.
Passing to the vector notation
\be\label{vector}
{\widetilde F}{}^{AB} =\frac{1}{2} \epsilon^{ABCD} F_{CD}, \quad {\widetilde W}{}^{AB} =\frac{1}{2} \epsilon^{ABCD} W_{CD}, \quad
\ee
where
\be\label{vec1}
{\widetilde F}{}^{AB} = \frac{1}{2}\Big[ \big( \sigma^{AB}\big)^{\alpha\beta} F_{\alpha\beta} + \big( {\tilde\sigma}{}^{AB}\big)^{\dot\alpha\dot\beta} \bF_{\dot\alpha\dot\beta} \Big]
\ee
and
\be\label{vec2}
{\widetilde W}{}^{AB} = \frac{1}{2}\Big[ \big( \sigma^{AB}\big)^{\alpha\beta} \frac{1 + \frac{1}{2}\bV^2}{1 - \frac{1}{4}V^2 \bV^2} V_{\alpha\beta}
+ \big( {\tilde\sigma}{}^{AB}\big)^{\dot\alpha\dot\beta} \frac{1 + \frac{1}{2}V^2}{1 - \frac{1}{4}V^2 \bV^2}\bV_{\dot\alpha\dot\beta} \Big],
\ee
one may rewrite the relations \p{Fphys} as
\be\label{Fvec}
{\widetilde F}{}^{AB} = \det \cE \big(\cE{}^{-1}\big)^{A}_C \big(\cE{}^{-1}\big)^{B}_D \; {\widetilde W}{}^{CD}.
\ee
Note that, being written in terms of $W^{AB}$ \p{vec2}, the action \p{act2} acquires the familiar form of the Born-Infeld action
\be\label{BI1}
S= \int d^4 x \det \cE \bigg[ 1+ \sqrt{1+\frac{1}{2}W_{AB} W^{AB} -\frac{1}{16} \big( W_{AB}{\widetilde W}{}^{AB}\big)^2}\; \bigg],
\ee
and thus it gives the component action of the famous $N=1, d=4$ supersymmetric Born-Infeld theory \cite{CF}.
\subsection{Complete proof of the invariance with respect to unbroken supersymmetry}
Now we are ready to prove the invariance of the supersymmetric Born-Infeld action \p{act2} under unbroken supersymmetry. Substituting the function $G$ \p{G}
into the variation \p{Qsusylagr} we will get
\be\label{Qsusylagr2}
\delta_Q {\cal L} = -2\im\, \det \cE \epsilon^{\alpha} \cD_{\beta\dot\alpha} \bpsi^{\dot\alpha} V_{\alpha}^{\beta}
\frac{1 + \frac{1}{2}\bV^2}{1 - \frac{1}{4}V^2 \bV^2}
+ 2\im\, \epsilon^{\alpha}\cD_{\alpha\dot\beta}\bpsi^{\dot\alpha} \bV^{\dot\beta}_{\dot\alpha} \frac{1 + \frac{1}{2}V^2 }{1 - \frac{1}{4}V^2\bV^2 }\,.
\ee
Integrating by parts and discarding total derivatives, one finds that
\be\label{Qsusylagr3}
\delta {\cal L} = 2\im\, \epsilon^{\alpha}\bpsi^{\dot\alpha} \partial_{\gamma\dot\gamma}\bigg[ \det \cE\, \big(\cE^{-1}\big)_{\beta\dot\alpha}^{\gamma\dot\gamma}\, V_{\alpha}^{\beta}
\frac{1 + \frac{1}{2}\bV^2}{1 - \frac{1}{4}V^2 \bV^2} - \det\cE\, \big(\cE^{-1}\big)_{\alpha\dot\beta}^{\gamma\dot\gamma}\,
\bV^{\dot\beta}_{\dot\alpha} \frac{1 + \frac{1}{2}V^2 }{1 - \frac{1}{4}V^2\bV^2 } \bigg].
\ee
Remarkably, the variation of the Lagrangian is not proportional to the Bianchi identity \p{BI3}, as one may expect.
However, using the Bianchi identity \p{BI3} this variation \p{Qsusylagr3} can be simplified to be
\be\label{Qsusylagr4}
\delta {\cal L} = -2 \det\cE\, V^{\sigma}_{\rho} \cD_{\sigma\dot\gamma} \big(\epsilon^{\alpha}\psi_{\alpha} \big)\, \cD^{\rho\dot\gamma}\big(\bpsi^{\dot\alpha}\bpsi_{\dot\alpha}   \big)
\frac{1 + \frac{1}{2}\bV^2}{1 - \frac{1}{4}V^2 \bV^2}
+2 \det\cE\, \bV^{\dot\rho}_{\dot\sigma} \cD_{\rho\dot\rho} \big(\epsilon^{\alpha}\psi_{\alpha} \big)\, \cD^{\rho\dot\sigma}\big(\bpsi^{\dot\alpha}\bpsi_{\dot\alpha}   \big)
\frac{1 + \frac{1}{2}V^2 }{1 - \frac{1}{4}V^2\bV^2 }\,.
\ee
Once again, integrating by parts to remove the derivative from $(\epsilon^{\alpha}\psi_{\alpha})$ and using once more the identity \p{BI3}, we will obtain
\bea\label{Qsusylagr5}
\delta {\cal L} &=& 4\im \det\cE\, \big(\epsilon^{\alpha}\psi_{\alpha} \big)\, \cD^{\lambda\dot\lambda} \big(\bpsi^{\dot\alpha}\bpsi_{\dot\alpha}   \big)
\bigg[ \frac{1 + \frac{1}{2}\bV^2}{1 - \frac{1}{4}V^2 \bV^2}\, V^{\sigma\rho} \cD_{\rho\dot\gamma}\psi_{\lambda} \cD_{\sigma}^{\dot\gamma}\bpsi_{\dot\lambda}
-    \frac{1 + \frac{1}{2}V^2}{1 - \frac{1}{4}V^2 \bV^2}\, \bV^{\dot\sigma\dot\rho} \cD_{\gamma\dot\sigma}\psi_{\lambda} \cD^\gamma_{\dot\rho} \bpsi_{\dot\lambda} \bigg] +\nn \\
&& +2\det\cE\, \big(\epsilon^{\alpha}\psi_{\alpha} \big) \bigg[ V^\sigma_\rho \cD_{\sigma\dot\gamma} \cD^{\rho\dot\gamma} \big(\bpsi^{\dot\alpha}\bpsi_{\dot\alpha}   \big)
\frac{1 + \frac{1}{2}\bV^2}{1 - \frac{1}{4}V^2 \bV^2} - \bV^{\dot\rho}_{\dot\sigma} \cD_{\gamma\dot\rho}\cD^{\gamma\dot\sigma}
\big(\bpsi^{\dot\alpha}\bpsi_{\dot\alpha}   \big)   \frac{1 + \frac{1}{2}V^2}{1 - \frac{1}{4}V^2 \bV^2}     \bigg].
\eea
Finally, replacing in the second line the double derivatives acting on $\big(\bpsi^{\dot\alpha}\bpsi_{\dot\alpha}   \big)$ by the commutator
$\cD_{(\sigma\dot\gamma} \cD^{\dot\gamma}_{\rho)} = \frac{1}{2} \big[ \cD_{(\sigma\dot\gamma}, \cD^{\dot\gamma}_{\rho)}   \big]$, and using the
expression for this commutator \p{nabladercomms} one may check that all terms in \p{Qsusylagr5}  cancel out.
Thus, the action \p{act2} is invariant with respect to both broken and unbroken supersymmetries.

\section{Conclusion}
In this paper, following the approach developed in \cite{BKS1,BKS2}, we derive the component on-shell action of the space-filling D3-brane, i.e. $N=1$
supersymmetric Born-Infeld action. In contrast with the cases of p-branes where irreducibility constraints can be immediately covariantized, the direct covariantization of the Bianchi identity within the standard scheme leads to  the equations of motion.  To overcome this problem, we introduced an additional Goldstone superfield associated with the generator of the automorphism transformation of the fermionic part of $N=2, d=4$ Poincar\'e superalgebra.  The first component of this superfield is the auxiliary field of the vector supermultiplet and thus, being in an off-shell situation, one may obtain the covariant  Bianchi identity and thus derive the component action. Of course, the on-shell component action for the $N=1$ Born-Infeld theory can be constructed from the superfield one \cite{CF, BG2,RT}. However, the action obtainable in such a way will contain a long tail of fermionic terms without any visible symmetry, whereas in the present approach all fermionic terms combine into covariant, with respect to broken supersymmetry, objects: covariant derivatives and vierbein. The D3-brane component action, being written in terms of these covariant objects,
has a very simple form. Similarly to the cases of p-branes, it mimics the bosonic Born-Infeld action.

In addition, the present results and the idea to introduce into the game some additional Goldstone superfields to go off-shell are the good start point to
re-consider $N=2,d=4$ Born-Infeld theory, with the hidden spontaneously broken $N=2$ supersymmetry \cite{N2BI1,N2BI2,N2BI3} within the nonlinear realization approach.

\section*{Acknowledgments}
The work of N.K. and S.K. was supported by RSCF grant 14-11-00598. The work of A.S. was partially supported by RFBR grant 15-52-05022 Arm-a.

\end{document}